\documentclass[aps,11pt]{revtex4}

\begin{document}

\tighten

\draft

\title{Decoherence of a Single Qubit from Quantum Anomaly}

\author{W.F. Chen, R. Kobes and G. Kunstatter}
\email{w3chen@sciborg.uwaterloo.ca, r.kobes@uwinnipeg.ca, g.
kunstatter@uwinnipeg.ca}
 \affiliation{Department of Physics, University of Winnipeg, Winnipeg,
Manitoba, Canada R3B 2E9 }

\begin{abstract}
\noindent We show via an explicit example that quantum mechanical
anomalies can lead to decoherence of a single quantum qubit through
phase relaxation. The anomaly causes the Hamiltonian to develop an
non-self-adjoint piece due to the non-invariance of the domain of
the Hamiltonian under symmetry transformation. The resulting
decoherence originates completely from the dynamics of the system
itself and not from interactions with the environment.

\end{abstract}

\maketitle

\vspace{3ex}

A physically realizable quantum computer must satisfy some delicate
requirements \cite{chuang}. One of these requirements is that
coherence  must be maintained within a single qubit and also among
entangled qubits. Coherence within a single qubit requires  dynamics
of the two-level quantum state to be controlled by unitary
evolution. This in turn is guaranteed by the self-adjointness of the
Hamiltonian in the Schr\"{o}dinger equation.  Up to now, attention
has been given mostly to decoherence that originates from the
interaction of the quantum system with its external environment
\cite{zurek}.  The purpose of this letter is to point out that
decoherenece can also come from anomalous symmetry breaking of the
quantum mechanical system. The novelty of this phenomenon is that
the decoherence originates from the system itself and not via
interactions do with an external environment. This anomalous
decoherence, which we  make explicit in the following via a toy
model, is potentially significant for quantum information theory and
should in principle be taken into account in the construction of
quantum computing models.

The model we consider is an electron in a magnetic field produced by
the Dirac monopole. It is described by the following Hamiltonian,
\begin{eqnarray}
H=\frac{[{\bf{\sigma}} \cdot ({\bf p}-e{\bf A})]^2}{2m} =
\frac{({\bf p}-e{\bf A})^2}{2m}-\frac{e}{2m} {\bf \sigma}\cdot {\bf
B} \label{hami}
\end{eqnarray}
where ${\bf A}$ is the singularity-free vector potential of the
Dirac magnetic monopole \cite{yang} and $B=\nabla \times {\bf
A}=g{\bf r}/{r^3}$, the corresponding magnetic field. Further, the
single-valueness of the wave function requires that $2eg$ should be
an integer.

 The model (\ref{hami}) possess a rotational symmetry
$SO(3)$ as well as a dynamical superconformal symmetry $OSP(1,1)$
\cite{dhoker}. The $SO(3)$ symmetry is generated by the angular
momentum of the electron-monopole system, ${\bf J}={\bf r}\times
({\bf p}-e{\bf A})-eg {\bf r}/r+{\bf \sigma}/2$. The $OSP(1,1)$
consists of two parts: one is the  conformal symmetry $SO(2,1)$
generated by the Hamiltonian $H$, the dilatation operator $D$ and
the conformal generator $K$, and the other part is the $N=1/2$
conformal supersymmetry generated by the supercharge $Q$ and
conformal supersymmetry generator $S$. The $SO(2,1)$ conformal
symmetry is a generic feature of physical systems with $1/r^2$
potential \cite{jackiw}, whose algebra is realized as $[H,D]=iH,
~~[H,K]=2iD, ~~[D,K]=iK$.

The large symmetry described above allows the model (\ref{hami}) to
be solved exactly  with a suitable representation of $SO(3)\times
OSP(1,1)$ \cite{dhoker}. The quantum states are characterized by the
eigenstates $|j,m,\alpha,E\rangle$ of a complete set of compatible
operators $J^2$, $J_z$, $\mbox{sign}A$ and $H$, respectively. The
dynamical operator $A=\sigma \cdot \left({\bf J}+ eg {\bf
r}/{r}\right)-{1}/{2}$ is related to the Casimir of $OSP(1,1)$, and
the eigenvalues $\alpha=\pm 1$ of $\mbox{sign}A$ describe the two
helicity states of the electron related by the superconformal
transformation. Hence the state of the system is given by
\cite{dhoker}
\begin{eqnarray}
&& H |j,m,\alpha,E\rangle =E |j,m,\alpha,E\rangle \nonumber\\
&& J^2|j,m,\alpha,E\rangle =j(j+1) |j,m,\alpha,E\rangle,
 ~~~j=eg-\frac{1}{2}, eg+\frac{1}{2}, \cdots \nonumber\\
&& J_z|j,m,\alpha,E\rangle =m |j,m,\alpha,E\rangle,
~~~m=-j, -j+1, \cdots, j-1,j\nonumber\\
&& A|j,m,\alpha,E\rangle =\alpha d_j |j,m,\alpha,E\rangle,~~~
\alpha=\pm 1, ~~ d_j=\left[\left( j+\frac{1}{2}\right)^2-e^2g^2
\right]^{1/2}
\end{eqnarray}
 The wave function in spherical coordinate and the Pauli
two-component representation is $\Psi_E (r,\theta,\phi)=\langle
r,\theta,\phi,
\sigma|j,m,\alpha,E\rangle=\Phi_E(r)\eta_{j,m,\alpha}(\theta,\phi)$
\cite{dhoker}. The angular part $\eta_{j,m,\alpha}(\theta,\phi)$ can
be expressed explicitly in terms of the monopole harmonics
\cite{yang}.

Once the angular part of the wave function has been fixed, the
Hamiltonian (\ref{hami}) reduces to
\begin{eqnarray}
H = -\frac{1}{2m}\frac{1}{r} \frac{d^2}{dr^2}r+\frac{1}{2m r^2} (-A)
(-A+1). \label{redhami}
\end{eqnarray}
The radial eigenfunction $\Phi_{\rm E}(r)$ is the solution to the
eigenvalue equation $H\Phi_E(r)=E\Phi_E(r)$.

We are interested only in bound states ($E<0$) since the goal is to
describe a system which can be used for quantum computing. The bound
state radial eigenfunction reads
\begin{eqnarray}
\Phi_E(r)&=& N r^{-1/2}K_{2\delta_{j,\alpha}-1}(\beta r), \nonumber\\
\beta &=&(-2mE)^{1/2},
~~\delta_{j,\alpha}=\frac{1}{2}-\frac{1}{4}\alpha+\frac{1}{2} d_j
\end{eqnarray}
where $K_\nu $ is the modified Bessel function of the second kind
and $N$ is a normalization constant. It is easy to see that only
when $\nu <1$, $K_\nu$ is normalizable over the region containing
the origin. It turns out that there exists only one such bound
state, $\Phi_E(r)={2\beta K_{1/2}(\beta r)}/\sqrt{\pi r}$, which
arises when $j=|eg|-1/2$. In particular, the orbit angular momentum
$L^2$ is diagonal in the basis $|j=|eg|-1/2,m,\alpha,E\rangle$ and
there exists ${\bf \sigma}\cdot {\bf r}/{r}|j=|eg|-1/2, m\rangle=\pm
|j=|eg|-1/2, m\rangle$ \cite{dhoker}.

However, in this case  $\Phi_E(r)$ is singular at the origin, and we
need to perform regularization and renormalization operations on the
Hamiltonian (\ref{hami}) to make it regular. It should be noted that
the regularization of the model (\ref{hami}) is considerably
complicated. First, the solvability of the theory depends on the
larger dynamical symmetry $OSP(1,1)$, which should be preserved as
much as possible by the regularization scheme. Second, the theory
has a $U(1)$ gauge symmetry encoded in the angular wave function.
The regularization should keep the angular part
$\eta_{j,m,\alpha}(\theta,\phi)$ intact so that the Hamiltonian can
reduce to the form (\ref{redhami}). Otherwise, the exact solvability
of theory will be ruined. The regularization scheme we use is
described as follows. First, we observe that the reduced Hamiltonian
(\ref{redhami}) at $j=|eg|-1/2$ becomes
\begin{eqnarray}
H &=& -\frac{1}{2m}\frac{1}{r} \frac{d^2}{dr^2}r+\frac{1}{2m}
\frac{{\bf L}^2-e^2g^2-eg{\bf \sigma}\cdot {\bf e}_r}{r^2}
\nonumber\\
&=&-\frac{1}{2m}{\bf
\nabla}^2-\frac{e^2g^2}{2m}\left(1+\frac{1}{|eg|} \right)
\frac{1}{r^2} \label{reduce}
\end{eqnarray}
The Hamiltonian (\ref{reduce}) implies that at $j=|eg|-1/2$ the
radial sector of the spinning particle is equivalently described by
a spinless particle  in a spherically symmetric potential
$V(r)=-{\lambda}/{r^2}$, $\lambda\equiv \left(1+{1}/{|eg|} \right)
{e^2g^2}/{(2m)}$.

We choose a real-space cut-off regularization by introducing a
length scale $L$  as the regulator and re-defining the effective
potential as $V_R(r)=-{\lambda}/{r^2}\theta (r-L)$,
 $\theta$ denoting the Heaviside function.
The regularized energy eigenvalue equation
$\widetilde{H}\widetilde{\Phi}_E(r)={E}\widetilde{\Phi}_E (r)$ reads
\begin{eqnarray}
\left[\frac{d^2}{dr^2}+\frac{2}{r}\frac{d}{dr}-\frac{|eg|(|eg|+1)}{r^2}
+\frac{2m\lambda}{r^2}\theta (r-L) +2mE
\right]\widetilde{\Phi}_E(r)=0
\end{eqnarray}
The normalizable bound state solution expressed in the modified
Bessel functions is
\begin{eqnarray}
\widetilde{\Phi}_E(r) &=&  A r^{-1/2} I_{|eg|+1/2}(\beta r), ~~r <L,
\nonumber\\
&=& Br^{-1/2}K_{1/2}(\beta r), ~~~~~~ r>L \label{regwf}
\end{eqnarray}
The continuity of $\widetilde{\Phi}_E(r)$ at $r=L$ yields
$A=B{K_{1/2}(\beta L)}/{I_{|eg|+1/2}(\beta L)}$, and
 the normalization condition $\displaystyle \int_0^{\infty}dr
 r^2|\widetilde{\phi}_E(r)|^2=1$
fixes $B={2\beta}/{\sqrt{\pi}}$ as $L\rightarrow 0$. Finally, the
continuity of $d\widetilde{\phi}_E (r)/dr$ at $r=L$ leads to
\begin{eqnarray}
1+2 \beta L=-\beta L\, \frac{I_{|eg|-1/2}(\beta
L)+I_{|eg|+3/2}(\beta L)}{I_{|eg|+1/2}(\beta L)} \label{coneq}
\end{eqnarray}
Further, the expansion of $I_\nu (x)$ near $x=0$ gives the lowest
order reduction of (\ref{coneq}) at $L\rightarrow 0$, $\beta
L=-\left(|eg|+1\right)$. Hence we get the regularized bound state
energy ${E}=-{\left(|eg|+1\right)^2}/{(2m L^2)}$.

Obviously, the regulator dependent $E$ is divergent as $L\to 0$, and
the spectrum is unbounded from below in this limit. There are two
ways to cure this pathology. The first one is to adopt the viewpoint
of the Wilsonian effective field theory \cite{wilson}. We directly
take the regulator $L$ as the cut-off length scale $\Lambda$ and
 consider the regularized Hamiltonian
 $\widetilde{H}\equiv -{\nabla}^2/{(2m)}+V_R$
as an effective Hamiltonian above the length scale $\Lambda$. The
bound state energy at $r=\Lambda$ is
\begin{eqnarray}
E_B=-\frac{\left(|eg|+1\right)^2}{2m \Lambda^2} \label{bounden1}
\end{eqnarray}
The second one is the traditional approach of calculating the
one-particle irreducible (1PI) effective action and performing a
renormalization procedure as advocated in Ref.\,\cite{camb2}. At the
renormalization scale $r=\Lambda$, we make the subtraction by
splitting $E=E_B+E_{\rm div}$. In order to enforce the physical
requirement that the wave function should vanish at the origin,
which is needed for the self-adjointness of the Hamiltonian, we
introduce a counterterm to the $1/r^2$ potential. This counterterm
cancels the short-distance divergence $E_{\rm
div}=\left({1}/{\Lambda^2}
-{1}/{L^2}\right){\left(|eg|+1\right)^2}/{(2m)}$ in the regularized
energy $E$. Furthermore, as in field theory, the counterterm should
be absorbed into the redefinition of the coupling constant
$\lambda$. One particular challenge in the present context is that
the condition $2eg\in Z$ must be preserved for quantum mechanical
consistency.  A detailed analysis of this procedure will be
presented elsewhere \cite{chen}.

It is clear that both of the above approaches break the $SO(2,1)$
conformal symmetry due to the unavoidable presence of a length scale
$\Lambda$. This is a direct manifestation of the conformal anomaly
in this system, which has been shown in the modification of the
$SO(2,1)$ commutator algebra through a deformation of the
Hamiltonian by the anomaly operator:
\begin{eqnarray}
H\rightarrow H+ \widehat{\cal A},  ~~~\widehat{\cal A}\equiv
-i[H,D]_A \label{defha}
\end{eqnarray}
The resulting anomalous conformal algebra \cite{camb2,esteve} is
composed of $[H,D]=iH+[H,D]_A$, $[H,K]=2iD+2t[H,D]_A$ and
$[D,K]=iK+t^2[H,D]_A$. The Heisenberg equation further reveals the
conformal anomaly as the non-conservation of the conformal charges,
$dD/dt=\widehat{\cal A}$, ${dK}/{dt}=2t\widehat{\cal A}$.

An algebraic calculation of the first anomalous commutator shows
that the anomaly operator $\widehat{\cal A}$ is directly related to
the scaling behavior of the $1/r^2$ potential at the quantum level
\cite{camb2},
\begin{eqnarray}
\widehat{\cal A}\equiv -i[H,D]_A\equiv i[H,D]+H=\left(
1+\frac{1}{2}{\bf r}\cdot {\bf \nabla}\right)V(r) \label{anop}
\end{eqnarray}
We use the regularized wave function (\ref{regwf}) and the
regularized potential $V_R$ to explicitly evaluate expectation value
of the anomaly operator (\ref{anop}),
\begin{eqnarray}
{\cal A} &=& \langle \widehat{\cal A}\rangle =\langle V_{\rm
eff}(r)\rangle +\frac{1}{2}\langle {\bf
r}\cdot {\bf \nabla}V_R(r)\rangle \nonumber\\
&=& \lim_{L\to 0}\int_0^{\infty}dr r^2\left(
1+\frac{1}{2}r\frac{\partial}{\partial r}\right)V_R(r)
|\widetilde{\Phi}_E(r)|^2 \label{expectanop}
\end{eqnarray}
A straightforward calculation gives
\begin{eqnarray}
{\cal A}=\frac{e^2g^2\beta^2}{m}\left(1+\frac{1}{|eg|}\right) =
-2E_B e^2g^2\left(1+\frac{1}{|eg|}\right) \label{anomaly}
\end{eqnarray}

On the other hand, an alternative and elegant interpretation on the
origin of the anomaly in the Hamiltonian formalism has been
presented in \cite{esteve}, where it was demonstrated that the
anomaly is due to the fact that the symmetry generator does not
leave the domain of definition of the Hamiltonian invariant. By a
careful observation on the Heisenberg equation, it had been shown
\cite{esteve} that the anomaly arises as ${\cal A}=i \langle \Psi
(t)|\left( H^\dagger-H\right)G|\Psi (t)\rangle$, $G$ denoting a
certain symmetry generator operator which is $D$ for the scale
symmetry. This means that the anomaly operator can formally written
as
\begin{eqnarray}
\widehat{\cal A}= \left(H^\dagger-H\right)G =-{i}{\cal A}|\Psi
(t)\rangle \langle \Psi(t)| \label{nonad}
\end{eqnarray}
According to the rigorous definition of a self-adjoint operator
\cite{capri}, Eq.\,(\ref{nonad}) implies that the Hamiltonian has
always acquired a non-self-adjoint piece once its domain of
definition cannot be preserved by the symmetry transformation.

 The non-self-adjointness induced by the anomaly greatly modifies
the quantum dynamics of the system. In the Heisenberg picture, the
generator $G$ satisfies  a generalized Heisenberg equation
\cite{esteve},
\begin{eqnarray}
\frac{dG}{dt}=\frac{\partial G}{\partial
t}+i\left[H,G\right]+i\left(H^\dagger-H\right)G =\frac{\partial
G}{\partial t}+i[H,G]+i\widehat{\cal A} \label{mheiseq}
\end{eqnarray}
and it implies the following time-evolution of $G$,
\begin{eqnarray}
G(t)= \exp \left[i\int_0^t
ds\left(H^\dagger(s)-H(s)\right)\right]\,\exp\left[i\int_{0}^t ds H
(s)\right]G(0)\exp\left[-i\int_{0}^t ds H (s)\right] \label{hepi}
\end{eqnarray}
In  the Schr\"{o}dinger picture, we have the time-evolution in terms
of the modified Hamiltonian shown in Eq.\,(\ref{defha})
\begin{eqnarray}
|\Psi(t)\rangle =\exp\left[-i\int_0^t ds H(s)-\int_0^t\,ds {\cal
A}|\Psi (s)\rangle \langle \Psi (s)|\right]|\Psi (0)\rangle
 \label{scpi}
\end{eqnarray}
The formal integration solution (\ref{scpi}) for $|\Psi(t)\rangle$
shows that in the presence of the anomaly the quantum system
undergoes a non-unitary evolution resultant from the anomaly. This
is consistent with the fact that anomalous effects in a quantum
theory contribute only to the imaginary part of the quantum
effective action \cite{luis}.

Turning to the model at hand, we take $G$ to be the  generator $D$
of the scale symmetry. The conformal anomaly arises only for the
normalizable bound state $\Psi_E (r,\theta,\phi,\sigma)$ in the
$s$-wave sector, and
 originates from its radial part $\Phi_E(r)$. Therefore,
 Eq.\,(\ref{scpi}) tells that the time-evolution for this
specific stationary state  should be
\begin{eqnarray}
|\Phi_E(t)\rangle=  e^{-i(E-i{\cal A})t }|\Phi_E(0)\rangle
 \label{tevo}
\end{eqnarray}
where the energy and anomaly are provided by Eqs.\,(\ref{bounden1})
and (\ref{anomaly}), respectively. Note that the completeness
condition $\displaystyle\sum_E|\Phi_E\rangle \langle \Phi_E|={\bf
1}$ is used in deriving Eq.\,(\ref{tevo}). Although all the energy
eigenstates, including the scattering states, should be taken into
account in the the completeness condition, the anomaly only pertains
to the bound state and vanishes for all other eigenstates.  Thus we
effectively take $|\Phi_E\rangle \langle \Phi_E|={\bf 1}$ and get
Eq.\,(\ref{tevo}).

We now consider the electron-monopole system as a physical model for
quantum computing. The quantum state we manipulate is just the
normalizable bound state $\Psi_E (r,\theta,\phi, \sigma)$ at
$j=|eg|-1/2$, its two-level spin degrees of freedom playing the role
of a qubit:
\begin{eqnarray}
\Psi_E (r,\theta,\phi, \sigma)&=&
\Phi_E(r)\eta_{j,m,\alpha}(\theta,\phi,\alpha) \equiv
f_1(r,\theta,\phi)\left(\begin{array}{c} 1 \\0
\end{array}\right)
+f_2(r,\theta,\phi)\left(\begin{array}{c} 0 \\1
\end{array}\right)
\label{qubit}
\end{eqnarray}
The spatial amplitudes $f_i(r,\theta,\phi)$ ($i=1,2$) can be
obtained with some algebraic operations \cite{dhoker}.

As we will show below, it is the time-evolution of
$f_i(r,\theta,\phi)$ related to $\Phi_E(r)$ that brings about the
decoherence between two spin states during a quantum computing due
to the presence of anomaly. Roughly speaking, the two spin states
constitutes a qubit, and one must control their dynamical evolution
to carry out information processing. We therefore switch on a
time-dependent Hamiltonian to make the spin flips that can
ultimately be used in a quantum algorithm. However, the spatial
sector $f_i(r,\theta,\phi)$ of the wave function  will evolves in
time controlled by the quantum effective Hamiltonian of the system
itself along with the spin flipping  dominated by the external
Hamiltonian. According to Eq.\,(\ref{tevo}), the anomaly will cause
$f_i(r,\theta,\phi,t)$ to have a damping factor which in turn will
lead to decoherence. In the following we show the details of how
this phenomenon happens.

Let us first analyze the quantum effective Hamiltonian provided by
the system itself. Obviously, the time-evolution (\ref{tevo}) of
$\Phi_E(r)$ of the bound state wave function gives the spatial part,
$H_{\rm spa}=E-i{\cal A}$. As for the spin sector, we use the fact
that at $j=|eg|-1/2$ the orbit- and spin- angular momenta decouple,
and the spin part of the wave function is the eigenstate of the
operator ${\bf \sigma}\cdot {\bf r}/r$. Specifically, the form of
the radial Hamiltonian (\ref{reduce}) shapes only when the
eigenvalue equation of the operator ${\bf \sigma}\cdot {\bf r}/r$
has been applied. So we can simply choose $H_{\rm spin}={\bf
\sigma}\cdot {\bf r}/r$. A combination of the spatial and spin
sectors determines that  the effective Hamiltonian with resect to
the bound state (\ref{qubit}) should take the following form:
\begin{eqnarray}
H_{\rm sys} &=& H_{\rm spa}\otimes H_{\rm spin} = \left(E-i{\cal
A}\right)\frac{{\bf \sigma}\cdot{\bf r}}{r} \nonumber\\
&=& (E-i{\cal A})\left(\sigma_x\sin\vartheta\cos\varphi
+\sigma_y\sin\vartheta\sin\varphi+\sigma_z\cos\vartheta\right)
\label{spha}
\end{eqnarray}
where $(\vartheta$,$\varphi$) represents the spin orientation in
three-dimensional space.

 Eq.\,(\ref{spha}) is the effective
Hamiltonian realized on the bound sate of the system. We now switch
on a time-dependent external Hamiltonian to make the spin flip. A
typical choice is the interaction of the spin with an oscillating
external magnetic field in two-dimensional $x-y$ plane, ${\bf
B}_{\rm ext} = B_0 \left(\cos\omega t {\bf e}_x+\sin\omega t {\bf
e}_x\right)$, and the Hamiltonian $H_{\rm ext}={e}/{2m}{\bf
\sigma}\cdot {\bf B}_{\rm ext}$. The spin dynamics is dominated by
the Schr\"{o}dinger equation
\begin{eqnarray}
 i\frac{\partial\Psi_E (t)}{\partial t} &=&  \left(H_{\rm
sys}+H_{\rm ext}\right)\Psi_E(t) =
\left\{\left[\frac{eB_0}{2m}\cos\omega t+\left(E-i{\cal
A}\right)\sin\vartheta\cos\varphi\right]\sigma_x \right.\nonumber\\
&& \left.+\left[ \frac{eB_0}{2m}\sin\omega t+\left(E-i{\cal
A}\right)\sin\vartheta\sin\varphi \right]\sigma_y +\left(E-i{\cal
A}\right)\cos\vartheta \sigma_z\right\}\Psi_E(t)
\end{eqnarray}
We neglect the $E-i{\cal A}$ term in the $\sigma_x$ and $\sigma_y$
components since usually the  microscopic values of the energy $E$
and the anomaly are much smaller than the macroscopic magnetic
field, $|E|,|{\cal A}| \ll|e|B_0/2m$. In this approximation
 the time-evolution of the spin state reads
\begin{eqnarray}
\Psi_E(t)=\exp\left\{-{\cal A}\cos\vartheta\sigma_z t-i\left[\left(
E\cos\vartheta-\frac{\omega}{2}\right)\sigma_z+\frac{eB_0}{2m}\sigma_x
\right]t\right\} \Psi_E (0) \label{solu2}
\end{eqnarray}

To show explicitly the occurrence of the decoherence implied from
$\Psi_E(t)$, we take $E\cos\vartheta=\omega/2$ as in nuclear
magnetic resonance and use again $|{\cal A}| \ll|e|B_0/2m$. Assume
that the initial state is spin-up, $\Psi_E
(0)=f_1(r,\theta,\phi)\left(1,0\right)^{T}$,  Eq.\,(\ref{solu2})
yields
\begin{eqnarray}
\Psi_E(t)&=& c_1 (t)\left(\begin{array}{c} 1 \\ 0
\end{array}\right) +c_2(t)\left(\begin{array}{c} 0 \\ 1
\end{array}\right),
\nonumber\\
c_1(t) &=& \cos \left[\left(\frac{e^2B^2_0}{4m^2}-{\cal
A}^2\cos^2\vartheta\right)^{1/2}t\right] -{\cal
A}\cos\vartheta\,\frac{\displaystyle
\sin\left[\left(\frac{e^2B^2_0}{4m^2}-{\cal
A}^2\cos^2\vartheta\right)^{1/2}t\right]}{
\left(\displaystyle\frac{e^2B^2_0}{4m^2}-{\cal
A}^2\cos^2\vartheta\right)^{1/2}}
\nonumber\\
c_2(t) &=& e^{i\pi/2}\frac{eB_0}{2m}\,\frac{\displaystyle
\sin\left[\left(\frac{e^2B^2_0}{4m^2}-{\cal
A}^2\cos^2\vartheta\right)^{1/2}t\right]}{
\left(\displaystyle\frac{e^2B^2_0}{4m^2}-{\cal
A}^2\cos^2\vartheta\right)^{1/2}}
\end{eqnarray}
Clearly, the non-vanishing ${\cal A}$ leads to
$|c_1(t)|^2+|c_2(t)|^2\neq 1$, and hence the decoherence between the
two helicity states occurs and the qubit is destroyed.

To summarize, we have used an electron-monopole system to reveal a
phenomenon not previously discussed in the quantum computing
literature: a quantum mechanical anomaly can result in decoherence.
Note that anomaly is a quantum dynamical phenomenon rooted within
the system itself. It reflects how quantum effects can render a
classically feasible symmetry unrealizable. One typical example is
the case where the configuration space has non-trivial topology so
that the Hilbert space constructed via the quantization procedure
from the classical phase space cannot sustain all the classical
symmetries. In the case we have just considered, the source of the
anomaly is the singular behaviors of the interaction potential
 near the magnetic monopole. The
classical conformal symmetry does not preserve the Hilbert space as
the domain of definition of the Hamiltonian due to the singular
behavior of the wave function in the $s$-wave sector.

 Until now the search for a physically realizable quantum computer
 has been concerned only with decoherence that arises due to interactions
 with the external environment. It is important to emphasize
that decoherence can also in principle be induced by quantum
anomalies. Since this dissipation originates from the dynamics of
the quantum system itself, it seems that it has the potential of
being more destructive than the standard mechanisms for decoherence.

\vspace{2mm}

\noindent This work is supported by the Natural Sciences and
 Engineering Research Council of Canada.

\end{document}